\documentclass[twocolumn,showpacs,preprintnumbers]{revtex4}
\usepackage{graphicx}
\usepackage{dcolumn}
\usepackage{bm}

\def\be{\begin{equation}}
\def\ee{\end{equation}}
\def\bea{\begin{eqnarray}}
\def\eea{\end{eqnarray}}

\begin{document}

\title{Bose Einstein Condensation as Dark Energy and Dark Matter}
\author{Masako Nishiyama, Masa-aki Morita, and Masahiro Morikawa}
\date{\today}

\begin{abstract}
We study a cosmological model in which the boson dark matter gradually
condensates into dark energy. Negative pressure associated with the
condensate yields the accelerated expansion of the Universe and the rapid
collapse of the smallest scale fluctuations into many black holes, which
become the seeds of the first galaxies. The cycle of gradual sedimentation
and rapid collapse of condensate repeats many times and self-regularizes the
ratio of dark energy and dark matter to be order one.
\end{abstract}

\maketitle

%\subtitle{---Quantum Structure in Universe---}

\affiliation{Department of Physics, Ochanomizu University, 
2-1-1 Ohtuka, Bunkyo, Tokyo,112-8610 Japan
}

\textit{Introduction}

Recent observations including WMAP for cosmic microwave background \cite
{map:1} and supernovae Type Ia Hubble diagrams\cite{Riess:1999}
independently indicate that some unknown dark energy (DE) component
dominates the global cosmic energy density and induces the accelerated
expansion of the Universe. On the other hand, previous many observations
manifestly indicate that some unknown dark matter (DM) dominates the local
energy density, whose total abundance turns out to be the same order as that
of DE.

Amazingly without explicitly specifying dominant contents of the Universe,
the standard $\Lambda $CDM model can explain many of the observational data,
especially those related with the density fluctuations in the linear stage.
In the non-linear stage however, the standard $\Lambda $CDM model cannot
explain the very early formation of astronomical objects and re-ionization
at around $z\approx 20,$ which is observationally required. Some extra
mechanism such as biasing is inevitable for the galaxy formation, since the
gravity itself cannot sufficiently compact matter.

In this letter, we propose a unified model of DE and DM in the context of a
cosmic phase transition and further the mechanism for the early formation of
non-linear objects. The key feature of DE would be the volume-independent
negative pressure, which guarantees the accelerated cosmic expansion through
the Einstein equation. The only known matter which shows negative pressure
would be some classical coherent objects such as scalar fields. As an origin
of such coherent object, we specifically consider the Bose-Einstein
condensation (BEC) of a boson field with attractive interaction, and we can
identify the condensation as DE and the exited bose gas as DM. The pressure
of BEC for ideal gas is volume independent and therefore an introduction of
attractive interaction would easily yields negative pressure. If we adopt
this scenario, the DE dominance means that our universe is in almost the
ground state described by a macroscopic wave function. Thus the largest
scale of the Universe, as well as the smallest, would turn out to be
described by quantum mechanics.

Basic model we construct would have the following general features: (a)Bose
gas is introduced as DM which initially dominates the energy density and the
condensate of the boson is identified as DE. (b)The condensate has negative
pressure due to its potential which reflects the attractive interaction. For
the spatially uniform component of BEC, this negative pressure works as a
cosmological constant and guarantees the accelerated expansion. (c)The
sedimentation of the condensate slowly proceeds in the cosmic evolution.
(d)When the energy density of BEC exceeds some critical value, the
condensates rapidly collapses into compact boson stars and black holes,
which work as the standard cold dark matter and also become the seeds of
galaxies. Simultaneously a new sedimentation process begins. This cycle of
sedimentation-and- collapse repeats many times. These rapid collapses take
place well after the photon-decoupling stage, and therefore the large scale
structure predicted by the $\Lambda $CDM model and actually observed pattern
in CMB fluctuations would not strongly be violated.

\textit{cosmic BEC mechanism}

The BEC occurs when the thermal de Broglie length $\lambda _{dB}\equiv
\left( {2\pi \hbar ^{2}/\left( {mkT}\right) }\right) ^{1/2}$ exceeds the
mean separation length of particles $n^{-1/3}$, where $n=N/V$ is the mean
number density of the boson and $m$ is the boson mass. The critical
temperature is given by 
\begin{equation}
T_{C}=\frac{2\pi \hbar ^{2}n^{2/3}}{m\zeta \left( {3/2}\right) }.
\label{eq1}
\end{equation}
At the transition point, the chemical potential $\mu $ associated with the
conservation of particle number $N$ within the volume $V$ shows singular
behavior $\mu \rightarrow -0$. The condensation is possible only for
non-relativistic stage $T<m$ and the particle number $N$ is conserved. On
this stage, the cosmic energy density behaves as 
\begin{equation}
n=n_{0}\left( {\frac{m}{2\pi \hbar ^{2}}\frac{T}{T_{0}}}\right) ^{3/2},
\label{eq2}
\end{equation}
\noindent where $n_{0},T_{0}$ are the number density and the temperature at
some moment in the non-relativistic stage. In deriving eq.(\ref{eq2}), we
have assumed the adiabatic evolution in the sense that the entropy per
particle $s/n=\ln \left( {e^{5/2}\left( {mT}\right) ^{3/2}/\left( {2\pi
\hbar ^{2}}\right) ^{3/2}/n}\right) $ is conserved. The number density
dependence of the temperature $T\propto n^{2/3}$ is the same as Eq.(\ref{eq1}%
). Therefore the condition $T<T_{C}$ sets the upper limit of the boson mass;
if we adopt the value $\rho _{now}=9.44\times \mbox{ 
}10^{-30}\mbox{g / cm}^{\mbox{3}}$, then $m<2\mbox{ eV}$. Moreover, the
boson gas dominance condition in the non-relativistic stage sets the lower
limit of the mass $m>2\mbox{ eV}$, provided the boson field has the same
temperature as that of radiation. However, if the boson field is not the
thermal origin, there is no lower limit for the mass. \ If the above
conditions hold, the BEC process starts when the boson becomes
non-relativistic and it continues afterwards.

\textit{Quantum liquid model of cosmic BEC}

In order to describe the dynamics of BEC, the mean-field analysis\cite
{Meystre:2001} using the Gross-Pitaevskii equation is usually adopted. This
equation has a form of non-linear Schr\"{o}dinger equation, 
\begin{equation}
i\hbar \frac{\partial \psi }{\partial t}=-\frac{\hbar ^{2}}{2m}\Delta \psi
+V\psi +g\left| \psi \right| ^{2}\psi ,  \label{eq3}
\end{equation}
\noindent where $\psi \left( {\vec{x},t}\right) $ is the condensate wave
function, $V\left( \vec{x}\right) $ is the potential, $g=4\pi \hbar ^{2}a/m$%
, and $a$ is the s-wave scattering length. If we decompose the wave function
as $\psi =\sqrt{n}e^{iS}$, and define the velocity as $\vec{v}=\hbar \vec{%
\nabla}S/m$, then Eq.(\ref{eq3}) reduces to the continuity equation, and the
hydrodynamic equation, 
\begin{equation}
m\frac{\partial \vec{v}}{\partial t}+\vec{\nabla}\left( {\frac{mv^{2}}{2}%
+V+gn-\frac{\hbar ^{2}}{2m\sqrt{n}}\Delta \sqrt{n}}\right) =0,  \label{eq4}
\end{equation}
\noindent except the last term which is quantum origin ($\propto \hbar ^{2})$%
. This term can be neglected if the wave number of the mode $k$ satisfies $%
k^{2}<k_{c}^{2}\equiv 2mgn/\hbar ^{2}$; i.e. large scale mode. Further, if
we choose the attractive interaction ($g<0)$, BEC can be described as fluid
with negative pressure. This type of argument on BEC yields the description
by general liquid with the equation of state $p=-A\rho ^{\alpha }$.

In this letter, we introduce the following simplest model for the BEC: Boson
gas is identified as cold DM with the equation of state $p=0,$ and the
condensate as DE with $p=-\rho $. The sedimentation of BEC in the uniform
expanding Universe slowly proceeds with the time scale $\Gamma ^{-1}$. This
setting is very similar to the chaplygin gas model\cite{Kamenshchik:2001}
with the equation of state $p=-A/\rho $, except that in the latter, DE and
DM properties are simultaneously included in this single equation of state
of a single phase.

Since the Universe is initially extremely uniform, this condensation would
also be uniform. The energy density of the excited bose gas is diluted by
the cosmic expansion, however, that of condensate is not diluted. This is
because the work supplied to expand the volume $V$ to $V+dV$ is $-pdV=\rho
dV $, which is exactly the necessary and sufficient amount of energy to
produce the new condensate in the region $dV$ with the same energy density.
Therefore eventually the condensate would dominate the excited gas component
and the expansion law of the Universe changes from the decelerated expansion
to the accelerated expansion.

In the early stage when the bose gas density dominates that of condensate,
density fluctuations are controlled by the dominant component, i.e. the bose
gas, and their evolution is described by the standard $\Lambda $CDM model.
However in the later stage when the condensate density dominates the bose
gas density, the situation drastically changes. The linear perturbation
equation for the gauge invariant quantity, in our case of the equation of
state $p=-\rho $, reduces to\noindent\ 
\begin{equation}
\delta _{k}^{\prime \prime }+5\delta _{k}^{\prime }=-\left( {6-\left( {\frac{%
k}{aH}}\right) ^{2}}\right) \delta _{k},  \label{eq8}
\end{equation}
where $\left( \cdots \right) ^{\prime }\equiv \frac{d\left( \cdots \right) }{%
d\ln a}$, and $\delta _{k}\equiv \delta \rho \left( k\right) /\rho .$
According to this, a small scale mode $\tilde{k}^{2}>6H^{2}$ rapidly grows,
and an almost cosmic horizon scale mode $\tilde{k}^{2}<6H^{2}$ slowly
decays, where the comoving wave number $\tilde{k}\equiv k/a$ is defined.
Moreover, the smaller the fluctuation scale, the faster the growing process: 
$\delta _{k}\propto \exp \left( {t\tilde{k}/H}\right) $. Note that this
rapid collapse is related with the fact that in the gas of negative
pressure, there is no sound wave ($c_{S}^{2}<0)$. The situation now
considering is not the (never growing) density fluctuations in the de Sitter
space in which the pressure is a strict constant.

Let us consider the non-linear stage of the condensate collapse. We suppose
a uniform spherically distributed over-density region of BEC of the radius $%
r $ and the density $\rho $. Since the pressure gradient only exists on the
surface of the sphere, the surface is isotropically compressed to form a
dense skin. In this process, the skin region with the width $dr$ acquires
energy $4\pi r^{2}dr\left( {-p}\right) =4\pi r^{2}dr\rho $ which is exactly
the mass of the skin. Therefore the skin soon acquires the light velocity.
The skin itself has large negative pressure in magnitude and therefore
self-focuses. Since this skin is still located at the same pressure
gradient, it is further compressed and eventually wipe up the whole
condensate toward the center. This collapsing process can be expressed in
the evolution equation of the skin radius, 
\begin{equation}
\frac{d\left( {m_{t}\gamma \dot{r}}\right) }{dt}=-4\pi r_{t}^{2}\rho ,
\label{eq9}
\end{equation}
\noindent where $m_{t}=(4\pi /3)\left( {r_{0}^{3}-r_{t}^{3}}\right) \rho $
\noindent is the time dependent mass of the skin, and the right hand side of
Eq.(\ref{eq9}) is the total force acting on the skin $4\pi r_{t}^{2}p$. All
solutions of Eq.(\ref{eq9}) approach to the constant velocity solution 
\begin{equation}
\gamma =\sqrt{2},\;\dot{r}=1/\sqrt{2}.  \label{eq11}
\end{equation}
Thus the collapse of the condensate is almost the light velocity.

Self gravity of the sphere would further accelerate the skin collapse
especially in the later stage. The collapse would continue until the
Heisenberg uncertainty principle begins to support the structure, or a black
hole is formed, or it bounces back outward if the condensate melts at the
final stage of the collapse. Anyway the collapsed condensate forms localized
compact objects classified as cold dark matter.

\textit{Gradual sedimentation of the condensation--- Self Organized
Criticality}

We now turn our attention to the global evolution of DE/DM in the expanding
Universe. The evolution of the various energy densities are governed by the
set of equations, 
\begin{equation}
\begin{array}{l}
\rho =\rho _{c}+\rho _{g}+\rho _{l},\quad H\equiv \frac{\dot{a}}{a}=\sqrt{%
\frac{8\pi G\rho }{3}}, \\ 
\dot{\rho}_{c}=\Gamma \rho _{g},\quad \dot{\rho}_{g}=-3H\rho _{g}-\Gamma
\rho _{g},\quad \dot{\rho}_{l}=-3H\rho _{l},
\end{array}
\label{eq12}
\end{equation}
\noindent where $\rho _{c}$, $\rho _{g}$, and $\rho _{l}$ are the energy
densities of condensate, excited boson gas, and the localized energy density
after the rapid collapse, respectively.

This set of evolution equations are valid when the condensate does not
dominate the energy density: $\rho _{c}<\rho _{g}+\rho _{l}$. Once it
dominates $\rho _{c}>\rho _{g}+\rho _{l}$ after the time scale $\Gamma ^{-1}$
at around $z=z_{c}$, inhomogeneous components of the condensate would
rapidly collapse, and some fraction of $\rho _{c}$ is transformed into $\rho
_{l}$. Then the condition $\rho _{c}<\rho _{g}+\rho _{l}$ is recovered and
the gradual sedimentation of the condensate proceeds again following Eq.(\ref
{eq12}) during the time scale $\Gamma ^{-1}$. This repeated ``chase and
collapse'' process by DE($\rho _{c})$ and DM($\rho _{g}+\rho _{l})$
self-regularizes the ratio of them to fix order unity: $\rho _{c}\approx
\rho _{g}+\rho _{l}$. This kind of autonomous dynamics designated as Self
Organized Criticality (SOC) is widely known and observed in nature in
various places\cite{Maya:1}.

In the late stage when the Universe is locked in this SOC phase, let us
approximate $\rho _{c} = \rho _{g} + \rho _{l} $. Then the Einstein equation 
$\ddot {a}\left( t \right) = - \left( {4\pi G / 3} \right)\left( {\rho + 3p}
\right)a\left( t \right),\left( {\dot {a} / a} \right) ^{2} = \left( {8\pi G
/ 3} \right)\rho $ has the solution 
\begin{equation}  \label{eq13}
a\left( t \right) \propto t ^{4 / 3} ,\quad \rho \left( t \right) \propto
a\left( t \right) ^{ - 3 / 2}
\end{equation}
\noindent which corresponds to the deceleration parameter $q = - \ddot
{a}a\dot {a} ^{ - 2} = - 1 / 4$.

\begin{figure}[htbp]
\centerline{\includegraphics[width=8cm]{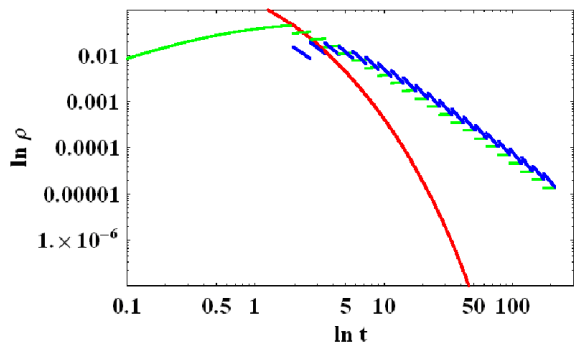}}
\caption{Time evolution of various densities. Green=$\protect\rho _{c} $,
Red=$\protect\rho _{g} $, and Blue=$\protect\rho _{l} $. We set the
parameter $\Gamma = 0.1$ in the unit of $8\protect\pi G / 3 = 1$, and the
transition rate from $\protect\rho _{c} $ to $\protect\rho _{l} $ is set to
be $1 / 3$. After 22 cycles of $\protect\rho _{c} $ dominance, the scale
factor has evolved about 200 times. DE($\protect\rho _{c} )$ and DM($\protect%
\rho _{g} + \protect\rho _{l} )$ self-regularize their ratio to be of order
unity. }
\label{fig1}
\end{figure}

\textit{Predictions and Observational tests of the model}

We now examine how our model is checked by observations.

\textbf{(a) Power spectrum of the density fluctuations in the linear regime }%
The collapse of the condensate proceeds in the smallest scale. This is
because the density fluctuation is stronger in the smaller scale, and the
collapse proceeds with almost the light speed. Therefore the BEC model does
not affect the power spectrum in the linear (large scale) regime however it
extremely enhances the smaller non-linear regime. Especially the many black
hole formation from DE in the very early stage is the most prominent
characteristic of our model. Such black holes gradually cluster in much
larger scale along the standard scenario of $\Lambda $CDM model.

\textbf{(b) Power spectrum of CMB }The condensation is supposed to dominate
well after the decoupling time ($z_{c}<1000$ or $\Gamma ^{-1}>4\times 10^{5}$
year). Therefore the temperature inhomogeneity generated within the
decoupling era would not be changed from the standard $\Lambda $CDM model.
However the integrated Sacks-Wolfe effect, which originates from the
non-uniform gravitational potential after the decoupling time, has a chance
to modify the spectrum. In our BEC model, strong non-linear structure forms
very rapidly in the smallest fluctuation size, which is supposed to be well
outside of the linear regime.

\textbf{(c) Void-wall structure }The collapse of the condensate repeats many
times and the smallest scale fluctuations dominantly collapse in each
process. Although in the earlier collapsing process the smallest scale
fluctuations (ex. galaxy scale) dominantly forms non-linearity, in the later
collapsing process fairly larger scale fluctuations, much larger than the
super clusters, have a chance to dominantly form non-linearity. Moreover in
the earlier collapsing process almost spherically symmetric objects would be
preferably formed and therefore the non-linearity is almost point-like. On
the other hand in the later collapsing process, since the scale is large,
much irregular fluctuations would dominantly collapse and therefore the
non-linearity can be plain-like. This difference comes from the fact that
the collapse is triggered by the pressure which works on surface and not by
gravity which works on volume. This plain-like non-linearity may
characterize the largest scale structure of the Universe.

\textbf{(d) A first star harbors a boson star. }In our model, once the
condensate dominates the comic density, vary rapid collapses occur
especially in the smallest scale of density fluctuations. Associated with
this process, the collapsing object can easily fragment into many pieces
because the pressure is always negative. Therefore we can expect the large
amount of collapsing objects. For bosons, only the Heisenberg uncertainty
principle can support the structure against the collapse. This is the
quantum pressure expressed in the last term of Eq.(\ref{eq4}). This
structure is known as the boson star\cite{Kaup:1968}. The size $R$ of the
object is order of the compton wave length: $\lambda _{comton}=2\pi \hbar
/(mc)\approx 2R$. This size must be larger than the Schwarzschild radius $%
R>2GM/c^{2}$ for this object not to collapse into a black hole. These
equations yield the critical mass for the boson star. 
\begin{equation}
M_{critical}\approx m_{pl}^{2}/m\equiv M_{KAUP},  \label{eq14}
\end{equation}
\noindent only below which a structure can exist. For example, $m=10^{-5}%
\mbox{ eV}$ yields the critical mass abut the Earth mass: $%
M_{critical}=10^{-5}M_{\odot }=M_{\oplus }$. These compact structures may
have captured in the process of the first star formation. However even this
is the case, such seed boson star and the condensate should be melted into
boson gas in the high temperature stellar center, leaving no detectable
relic in principle.

\textbf{(e) A first galaxy harbors a giant black hole. }If the fluctuation
mass is larger than the critical mass $M>M_{critical}$, then the collapse
inevitably continues until a black hole is formed. In this process, since $%
p<0,\;dV<0$, no heating is expected thermodynamically. However the
gravitational energy released in this collapse would be $GM^{2}/R$, and if
this amount of energy is used to heat up the condensate, then the
temperature would be, from $NT\approx GM^{2}/R$ and $R\approx GM$, $T\approx
GMm/R\approx m$. Precisely at this point the boson becomes relativistic $%
T\approx m$ and the particle number no longer conserves. Then the chemical
potential is not well defined and the condensation melts into the ordinary
gas with positive pressure. Therefore the boson gas stops collapse and
violently expands outward. In this process, some fraction of the condensate
would form a central black hole and the rest of the condensate would melt
into the ordinary gas and expand outward. The gravitational potential of
this structure attracts baryon to form a cluster around the central black
hole. If the size is appropriate there forms a galaxy, which harbors a black
hole in the center and the boson gas and baryon in the outskirts.

What would happen for the expanding boson gas? The expansion of once melted
boson gas keeps the relation $T\approx GMm/R,\quad \rho \approx M/R^{3}$,
\noindent and therefore $T\propto \rho ^{1/3}$. Since this temperature is
still below the critical temperature, the boson gas would eventually
condense and re-collapse. Then the condensation melts and the boson expands
again. Apparently this bounce repeats multiple times with dissipation until
the gas thermalizes completely.

Since the strong non-linearity is formed in the very early stage at around $%
z=z_{c}$, no extra biasing process is necessary in our model. In this sense
our model is strongly bottom-up type. Early formed stars and galaxies should
be the origin of the re-ionization of the Universe around $z=z_{c}$.

\textbf{(f) Ejection of matter }The above bounces of the condensate need not
to be isotropic. Especially when the total angular momentum is not
vanishing, the bounce would preferably strong in the direction of the
angular momentum. This is because the centrifugal force naturally forms
dilute cone region in the direction of angular momentum. Therefore we
naturally expect, for each bounce, that a pair of blobs of the condensate is
ejected with relativistic velocity parallel to the angular momentum. The
blobs drag baryon which generally emits radiation. The ejection of blobs is
expected to takes place repeatedly many times.

If the ejection period is sufficiently small and the aligned blobs are
simultaneously visible, then this may be observed as two jets emanating from
the center of the galaxy toward the direction parallel to the angular
momentum. The essential characteristic of our model is that the created jets
are all discrete, i.e. sequence of high-speed blobs, and not simply a
continuous flow of plasma gas. Although at present we cannot quantitatively
describe such highly non-linear processes, we can naturally expect the firm
positive correlation among the mass of the central black hole, the
size of the galaxy, and the mean separation of the ejected blobs.

\textit{Summary and Further Developments}

We have developed a unified cosmological model of DE and DM using the
Bose-Einstein condensation of a light boson. Fast formation of compact
objects and the self organized critical Universe are realised.

In the context of BEC, there are many laboratory experiments using alkaline
atom gas, in which almost ideal BEC is realized\cite{Anglin:2002}.
Especially the collapse dynamics of BEC with negative pressure \cite
{Gerton:2001} \cite{Saito:2002} may be useful to develop our cosmological
BEC model. It would be interesting if the Gross-Pitaevskii equation
supplemented by the dissipative term, which represents the transition
between the condensate and excited gas, would lead to the quasi-periodic
oscillation of BEC collapse. We also have to consider the coherence
of BEC from various aspects\cite{Meystre:2001}. These issues are now under
investigation.

Acknowledgement

One of the authors (MM) would like to thank Kei-ichi Maeda, Shin Mineshige,
Hide-aki Mouri, Masatoshi Ohishi, Fumiaki Shibata, and Paul Steinhardt for
fruitful discussions and valuable comments.

\end{document}